# An empirical study of market risk factors for Bitcoin

Shubham Singh

New York University

**Abstract**

The study examines whether fama-french equity factors can effectively explain the idiosyncratic risk and return characteristics of Bitcoin. By incorporating Fama-french factors, the explanatory power of these factors on Bitcoin's excess returns over various moving average periods is tested through applications of several statistical methods. The analysis aims to determine if equity market factors are significant in explaining and modeling systemic risk in Bitcoin.

## INTRODUCTION

Bitcoin, the first and most prominent cryptocurrency, has firmly established its presence in the financial landscape. The approval of the first Bitcoin ETF in January 2024, followed by several others, has significantly enhanced its visibility and acceptance among mainstream investors. This surge in institutional interest shows Bitcoin's growing importance as a serious class of financial asset. However, despite its increasing popularity, the academic exploration of Bitcoin remains relatively sparse. Academic research remains underexplored in the context of Bitcoin and cryptocurrencies, with just 1032 papers published on Bitcoin in 2022[1], compared to 496.01 thousand papers on Artificial Intelligence in 2021[2].

The Fama-French three-factor model[3], a widely used model in risk analysis, has been extensively used to explain the returns of traditional assets by incorporating market risk premium, size, and value factors. This paper seeks to apply the Fama-French model to analyze Bitcoin's returns. Specifically, it aims to determine whether Bitcoin, an asset class often touted for its independence from traditional financial systems, is influenced by the same systemic risks that affect equities.

To achieve this, an analysis is conducted using different moving average periods—10-week, 20-week, and 52-week—to evaluate by reducing noise in the data. The primary research questions guiding this study are: Does the Fama-French three-factor model effectively explain Bitcoin's returns? How do the market risk, size, and value factors influence Bitcoin's returns? Do these influences vary significantly across different moving average periods?

By addressing these questions, this study tests the applicability of a well-established financial model to a novel asset class but also provides deeper insights into Bitcoin's risk-return profile. Understanding whether traditional market factors can explain Bitcoin's returns could have profound implications for investors and policymakers. It could help in better risk management, portfolio diversification strategies, and regulatory approaches. As Bitcoin continues to integrate into the broader financial system, such insights become increasingly valuable.

**LITERATURE REVIEW**

Previous studies have explored Bitcoin's volatility, correlation with other asset prices, however there's limited work on how factors affecting financial markets for other market products affect bitcoin prices.

Ji, Chang, Lan, Hsu, and Valverde conducted empirical research on the application of the Fama-French Three-Factor Model (FFTFM) and a Sentiment-Related Four-Factor Model in the Chinese blockchain industry. Their study aimed to understand the factors influencing stock returns in this rapidly evolving sector of financial technology (FinTech). The authors highlighted the significance of blockchain technology in China and its potential applications across various industries. They emphasized the importance of investor sentiment in influencing stock returns, particularly in a dynamic environment like the blockchain industry. By incorporating sentiment analysis into their models, the researchers aimed to enhance the explanatory power of the traditional FFTFM. Their findings suggested that while the FFTFM performed relatively well in explaining portfolio returns, the addition of a sentiment factor improved the model's explanatory power. They observed that the Chinese blockchain industry did not exhibit the size effect but showed evidence of a book-to-market ratio effect. Additionally, they found that investor sentiment played a significant role in influencing stock returns, with more positive sentiment correlating with higher returns. The study's insights contribute to the understanding of stock return dynamics in the Chinese blockchain industry and underscore the importance of considering investor sentiment in financial modeling.[4]

In their study on cryptocurrency pricing factors, Wang and Chong utilized the Fama-MacBeth approach to analyze various factors' explanatory power. They examined classical equity-based risk factors like size, momentum, and value to growth, alongside volatility and liquidity factors. The study found that while macro factors showed weak explanatory power, attention factors occasionally had an impact. Notably, the factor model constructed from significant factors explained a substantial portion of cryptocurrencies' excess returns. Additionally, the authors tested the factors using individual cryptocurrencies and portfolios, finding that factors like bid-ask spread and Roll's measure played significant roles in explaining excess returns. These findings contribute to understanding the pricing dynamics of cryptocurrencies and align with previous research recognizing the effectiveness of size and momentum factors.[5] In this work, The model is applied to Bitcoin but from a perspective of understanding its relation with broader markets.

**METHODOLOGY**

The Fama-French three-factor model is an extension of the Capital Asset Pricing Model (CAPM) that seeks to explain asset returns through multiple risk factors rather than just the market risk factor. The model was developed by Eugene Fama and Kenneth French in the early 1990s to provide a better understanding of asset prices by including factors that capture additional sources of risk and return that the CAPM does not.

The Fama-French three-factor model is specified as:

$$R_{Bitcoin} - R_f = \alpha + \beta(R_m - R_f) + \sigma SMB + hHML + kMOM$$

Where:

$R_{Bitcoin}$ = total returns from Bitcoin

$R_f$ = risk-free rate of return

$R_m$ = broader market returns

$SMB$ = Size factor

$HML$ = Value factor

$MOM$ = Momentum factor

1. **Factors**
    a. Market Risk Premium (Rm - Rf): This is the traditional market factor that measures the excess return of the market over the risk-free rate. β (market beta) represents the sensitivity of Bitcoin's returns to the returns of the overall market. A higher beta indicates that the asset is more volatile compared to the market.
    b. SMB (Small Minus Big): The SMB factor represents the size premium. It captures the return differential between small-cap stocks and large-cap stocks. It is calculated as the

average return on small-cap portfolios minus the average return on large-cap portfolios. σ (SMB coefficient) measures Bitcoin's sensitivity to the size factor. A positive σ indicates that Bitcoin behaves more like a small-cap stock.

c. HML (High Minus Low): The HML factor represents the value premium. It captures the return differential between value stocks and growth stocks. Value stocks are typically characterized by high book-to-market ratios, while growth stocks have low book-to-market ratios. h (HML coefficient) measures Bitcoin's sensitivity to the value factor. A positive h indicates that Bitcoin behaves more like a value stock.

d. MOM (Momentum Factor): The momentum factor captures the tendency of stocks that have performed well in the past to continue performing well in the future, and vice versa for poorly performing stocks. It is typically constructed by sorting stocks based on their past returns and forming portfolios that go long on winners and short on losers. *k* (MOM coefficient) measures Bitcoin's sensitivity to the momentum factor. A positive k indicates that Bitcoin tends to follow the momentum strategy.

2. **Data**

The weekly Bitcoin return data, adjusted for the risk-free rate (RF), spanning from September 2014 to March 2024, and factors for the same period are used. An Ordinary Least Squares (OLS) regression is fitted for Bitcoin's excess returns over different moving average periods (weekly, 10-week, 20-week, and 52-week) to estimate the coefficients and their statistical significance. Figures 1, 2, 3, and 4 show plots of weekly, 10 weekly, 20 weekly, and 52 weekly moving average factor data.

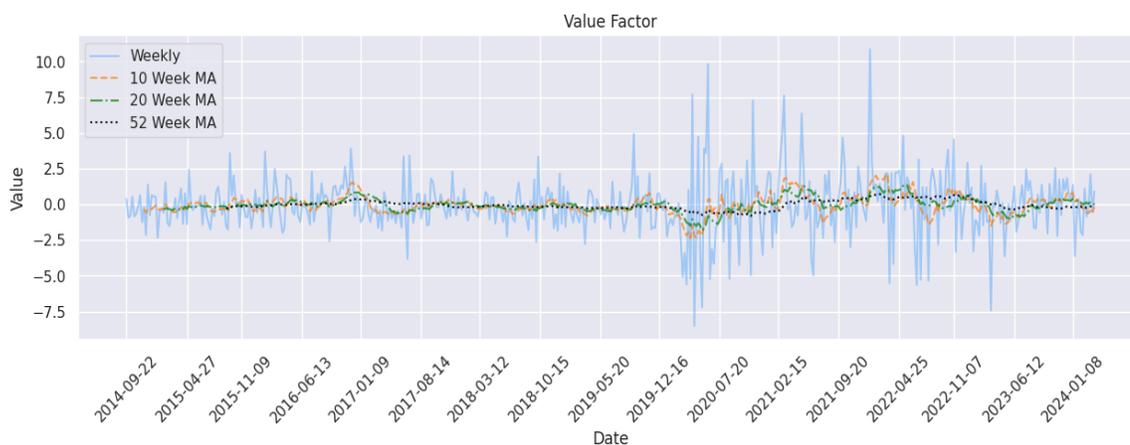

*Figure 1: Value factors over different moving averages*

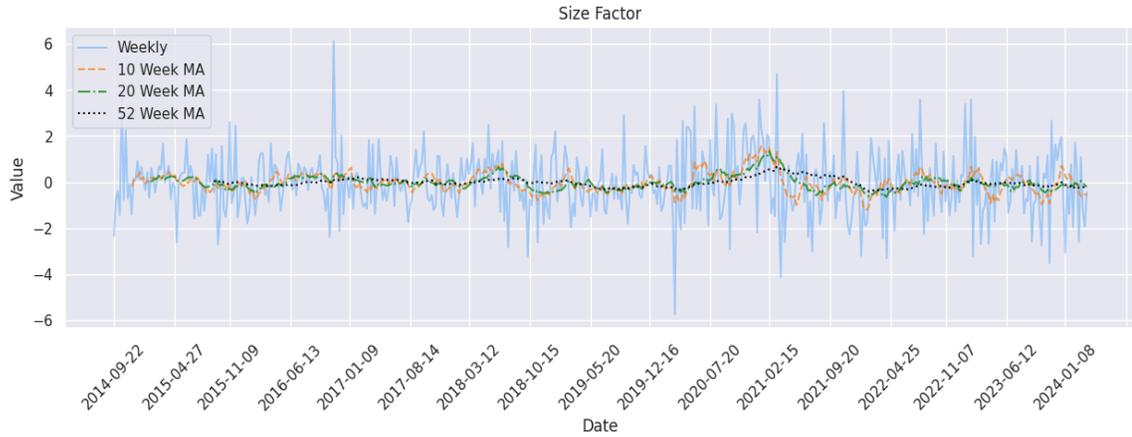

*Figure 2: Size factors over different moving averages*

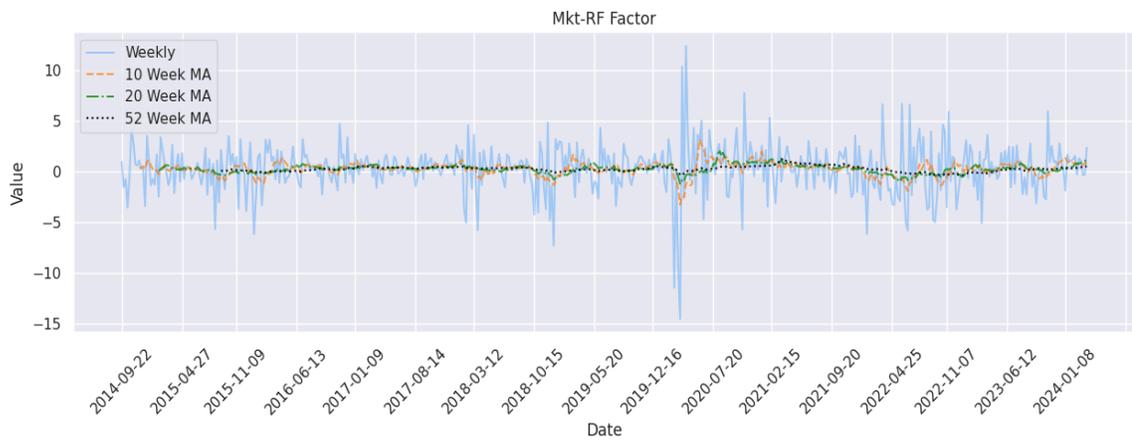

*Figure 3: Market return factors over different moving averages*

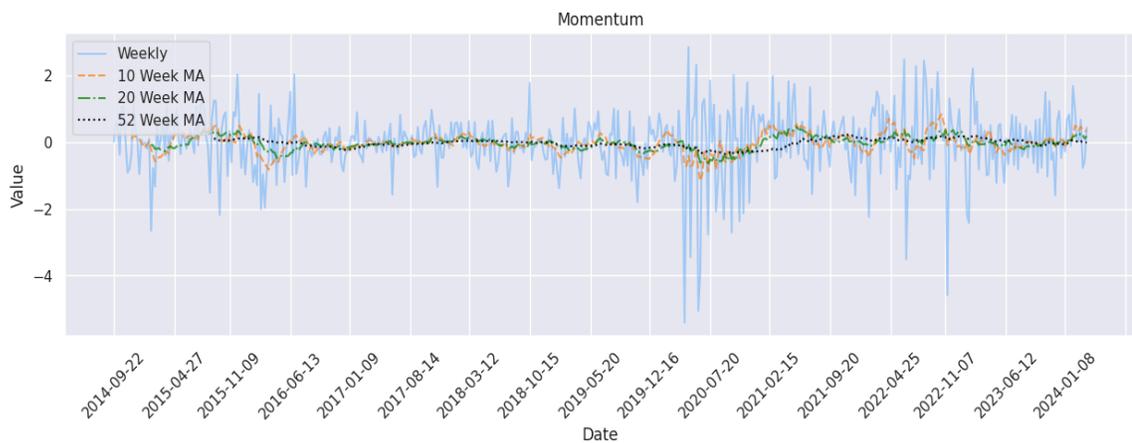

*Figure 4: Momentum factors over different moving averages*

**RESULTS**

The analysis yields coefficients for risk and returns in the Fama-French model for Bitcoin returns, the signals get better in quality as we reduce the noise with the use of longer moving averages. The R-squared values rise from a meager 0.005 for weekly data to 0.352 for the 52-week moving average, indicating a greater proportion of variance in Bitcoin returns explained by the model as the moving average period lengthens. Adjusted R-squared values show a similar trend, confirming the enhanced explanatory power. The F-statistics reinforce this, showing significant improvements in model fit, with p-values less than 0.0001 for all moving average models, whereas the weekly data model is not significant ($p = 0.682$).

The intercept is statistically significant across all models, with negative coefficients suggesting a consistent negative excess return not explained by the included factors. The market risk-free return is not significant in the weekly data model ($p = 0.340$) but becomes highly significant with moving averages ($p < 0.0001$), with increasing coefficients indicating a stronger relationship between market returns and Bitcoin returns over longer periods. The momentum factor is not significant in the weekly ($p = 0.579$) and 52-week ($p = 0.213$) models but shows moderate significance in the 10-week ($p = 0.022$) and 20-week ($p = 0.025$) moving averages, indicating a temporal impact of momentum on Bitcoin returns.

The size factor is not significant in the weekly data ($p = 0.547$) but is highly significant across all moving averages ($p < 0.0001$), with coefficients increasing over longer periods, suggesting a robust positive relationship between the size factor and Bitcoin returns. Similarly, the value factor is not significant in the weekly ($p = 0.597$) and 10-week ($p = 0.360$) models but approaches significance in the 20-week ($p = 0.065$) and is highly significant in the 52-week model ($p < 0.0001$), indicating a growing positive relationship between value stocks and Bitcoin returns over longer periods.

Diagnostic statistics indicate some issues with non-normality and autocorrelation, particularly in longer-term models. The Omnibus and Jarque-Bera tests show significant non-normality for the weekly and 52-week models ($p < 0.01$), while the 10-week model does not exhibit significant non-normality ($p = 0.438$). The Durbin-Watson statistics suggest potential autocorrelation issues, especially in models with moving averages, as values deviate from the ideal value of 2.

**DISCUSSION**

The 52-week moving average demonstrates the highest R-squared value, suggesting that traditional equity market factors are more effective in explaining Bitcoin's returns over longer periods. This enhanced explanatory power with longer moving averages highlights the need for further investigation into additional factors that may influence Bitcoin returns.

1. Market Risk Factor (Mkt-RF): The market risk factor consistently shows significant positive coefficients across all moving averages. This indicates a positive relationship with Bitcoin's returns, suggesting that Bitcoin's performance is influenced by overall market movements. The increasing significance and coefficient values with longer moving averages reinforce the idea that Bitcoin's returns are correlated with broader market trends over extended periods.
2. Size Factor (SMB): The size factor is significant across all moving averages, particularly in the 52-week model. This highlights the relevance of size effects in Bitcoin's performance, potentially reflecting the impact of liquidity and market capitalization on Bitcoin returns. The increasing significance of the size factor with longer moving averages suggests that Bitcoin, similar to smaller stocks, may benefit from higher returns associated with lower market capitalizations over longer time horizons.
3. Value Factor (HML): The value factor shows significance only in the 52-week moving average, suggesting that value effects are more pronounced over longer horizons. This indicates that Bitcoin behaves more like value stocks in the long run, aligning with the notion that assets with higher book-to-market ratios yield higher returns over extended periods.

Future research should consider additional factors that could explain Bitcoin returns, beyond the traditional Fama-French factors. Bitcoin's unique characteristics, such as its sensitivity to market sentiment and macroeconomic variables, warrant a more comprehensive analysis. Investigating sentiment-driven factors, such as social media trends, investor sentiment indices, and geopolitical events, could provide valuable insights into Bitcoin's return dynamics. This paper does not account for sentiment-driven influences, which are known to play a significant role in the cryptocurrency market.

**CONCLUSION**

This study extends the application of the Fama-French three-factor model to Bitcoin, revealing important insights into its risk-return profile. Our analysis shows that the model's explanatory power improves with longer moving averages, as evidenced by increasing R-squared values, with the 52-week moving average demonstrating the highest explanatory capability. The findings underscore the significance of market and size factors in explaining Bitcoin's returns, highlighting a consistently positive relationship with market returns and a notable impact of the size factor, particularly over longer periods. The value factor also shows significance in the long-term model, suggesting that Bitcoin exhibits characteristics similar to value stocks over extended horizons.

These results contribute to a better understanding of Bitcoin's financial dynamics, indicating that traditional equity market factors can partially explain Bitcoin's performance. However, the analysis also reveals the limitations of applying conventional models to Bitcoin without considering its unique attributes.

# APPENDIX

### Table 1: Significant statistics

| Statistic / Variable | Weekly data | 10-Week MA | 20-Week MA | 52-Week MA |
|---|---|---|---|---|
| R-squared | 0.005 | 0.131 | 0.184 | 0.352 |
| Adj. R-squared | -0.003 | 0.123 | 0.177 | 0.346 |
| F-statistic | 0.5735 | 18.14 | 26.61 | 59.95 |
| Prob (F-statistic) | 0.682 | 6.83e-14 | 6.36e-20 | 2.05e-40 |
| Log-Likelihood | 416.67 | 807.15 | 867.00 | 944.51 |
| AIC | -823.3 | -1604 | -1724 | -1879 |
| BIC | -802.3 | -1583 | -1703 | -1859 |
| Df Residuals | 492 | 483 | 473 | 441 |

**Coefficients and Statistics:**

### Table 2: Intercept related statistics

| Statistic | Weekly data | 10-Week MA | 20-Week MA | 52-Week MA |
|---|---|---|---|---|
| Return (coef) | -0.0133 | -0.0145 | -0.0154 | -0.0147 |
| Risk (std err) | 0.005 | 0.002 | 0.002 | 0.002 |
| t-value | -2.800 | -6.323 | -7.055 | -6.771 |
| p-value | 0.005 | < 0.0001 | < 0.0001 | < 0.0001 |

### Table 3: Market Risk-free Returns related statistics

| Statistic | Weekly data | 10-Week MA | 20-Week MA | 52-Week MA |
|---|---|---|---|---|
| Return (coef) | 0.0019 | 0.0145 | 0.0253 | 0.0357 |
| Risk (std err) | 0.002 | 0.004 | 0.005 | 0.007 |
| t-value | 0.956 | 3.827 | 4.720 | 4.853 |
| p-value | 0.340 | < 0.0001 | < 0.0001 | < 0.0001 |

### Table 4: Momentum related statistics

| Statistic | Weekly data | 10-Week MA | 20-Week MA | 52-Week MA |
|---|---|---|---|---|
| Return (coef) | -0.0027 | 0.0180 | 0.0234 | -0.0182 |
| Risk (std err) | 0.005 | 0.008 | 0.010 | 0.015 |
| t-value | -0.555 | 2.302 | 2.245 | -1.248 |
| p-value | 0.579 | 0.022 | 0.025 | 0.213 |

### Table 5: Size Factor related statistics

| Statistic | Weekly data | 10-Week MA | 20-Week MA | 52-Week MA |
|---|---|---|---|---|
| Return (coef) | 0.0021 | 0.0260 | 0.0352 | 0.0567 |
| Risk (std err) | 0.004 | 0.006 | 0.007 | 0.010 |
| t-value | 0.603 | 4.579 | 4.989 | 5.675 |
| p-value | 0.547 | < 0.0001 | < 0.0001 | < 0.0001 |

**Table 6: Value factor related statistics**

| Statistic | Weekly data | 10-Week MA | 20-Week MA | 52-Week MA |
|---|---|---|---|---|
| Return (coef) | 0.0012 | 0.0030 | 0.0073 | 0.0367 |
| Risk (std err) | 0.002 | 0.003 | 0.004 | 0.005 |
| t-value | 0.529 | 0.917 | 1.852 | 6.710 |
| p-value | 0.597 | 0.360 | 0.065 | < 0.0001 |

**Table 7: Omnibus, Durbin-Watson, Jarque-Bera:**

| Statistic | Weekly data | 10-Week MA | 20-Week MA | 52-Week MA |
|---|---|---|---|---|
| Omnibus | 11.726 | 1.650 | 19.473 | 10.002 |
| Prob(Omnibus) | 0.003 | 0.438 | 0.000 | 0.007 |
| Jarque-Bera (JB) | 16.691 | 1.664 | 17.348 | 10.370 |
| Prob(JB) | 0.000237 | 0.435 | 0.000171 | 0.00560 |
| Skew | 0.200 | -0.089 | -0.402 | -0.372 |
| Kurtosis | 3.803 | 2.777 | 2.525 | 2.928 |
| Durbin-Watson | 1.687 | 0.118 | 0.054 | 0.031 |
| Cond. No. | 2.56 | 3.91 | 6.05 | 11.2 |